\documentclass[12pt,twoside]{article}

\usepackage{jeep}       
\mysection{section}{\large \bf}{\thesection.~}   
\mysection{subsection}{\normalsize\bf}{\thesubsection.~}

\usepackage{a4wide}     

\usepackage{fancyheadings}

\usepackage{cite}   
\usepackage{epsf}
\usepackage{rotate}
\usepackage{subeqnarray}

\advance \headheight by 3.0truept       

\usepackage{times}

\newcommand{\lt}{\left}
\newcommand{\rt}{\right}
\newcommand{\ov}{\overline}
\newcommand{\eq}[1]{(\ref{#1})}
\def\openone{\leavevmode\hbox{\small1\kern-3.8pt\normalsize1}}%
\newcommand{\tw}{\ensuremath{\widetilde{\openone}}}
\newcommand{\1}{\ensuremath{\openone}}

\newcommand{\nn}{\nonumber \\}
\newcommand{\no}{\nonumber }
\newcommand{\fig}[1]{Fig.~\ref{#1}}
\newcommand{\tab}[1]{Tab.~\ref{#1}}
\newcommand{\bra}[1]{\langle \, #1 \, | }
\newcommand{\ket}[1]{| \, #1 \, \rangle }
\newcommand{\real}{{\rm Re}\,}
\def\lsim{{~\raise.15em\hbox{$<$}\kern-.85em
          \lower.35em\hbox{$\sim$}~}}

\newcommand{\prd}{Phys.~Rev.~D}
\newcommand{\plb}{Phys.~Lett.~B}
\newcommand{\npb}{Nucl.~Phys.~B}

\newlength{\miniwidth}
\newlength{\miniwidthplot}
\newlength{\nseparation}
\newenvironment{nfigure}[1]
        {\begin{figure}[#1]\hrule\vspace{\nseparation}\par}
        {\vspace{\nseparation}\par \hrule \end{figure}}
\newenvironment{ntable}[1]
        {\begin{table}[#1]\hrule\vspace{\nseparation}\par}
        {\vspace{\nseparation}\par \hrule \end{table}}

\begin{document}
~\\[-10truemm]
APCTP-97-17  \hfill DESY 97-120 \\
\phantom{APCTP-97-17  \hfill DESY 97-120 } \hfill  hep-ph/9710512 \\
~\vspace{2.43truecm}
\begin{center}
{\LARGE 
Probing penguin coefficients with the \\ lifetime ratio
$\tau \lt(B_s \rt)/\tau\lt(B_d\rt)$
 }\\[3\baselineskip]

\textsl{Yong-Yeon Keum\footnote{e-mail:keum@apctp.kaist.ac.kr},\\
APCTP, 207-43 Cheongryangri-dong, Dongdaemun-gu, 
Seoul 130-012, Korea, \\[2mm]
 and \\[2mm]
Ulrich Nierste\footnote{e-mail:nierste@mail.desy.de},\\
DESY - Theory group, Notkestrasse 85, D-22607 Hamburg, Germany,}\\[2mm]
\end{center}

\vfill
\begin{center}
\textbf{\large Abstract} 
\end{center}
We calculate penguin contributions to the lifetime splitting 
between the $B_s$ and the $B_d$ meson. In the Standard Model 
the penguin effects are found to be opposite in sign, but of 
similar magnitude as the contributions of the current-current 
operators, despite of the smallness of the penguin coefficients.  
We predict 
\begin{eqnarray}
\frac{\tau(B_s)}{\tau(B_d)} -1 &=& \lt( -1.2 \pm 10.0 \rt)
      \cdot 10^{-3} \cdot \lt(\frac{f_{B_s}}{190 \, \mathrm{MeV}} \rt)^2
, \no  
\end{eqnarray}
where the error stems from hadronic uncertainties.  Since penguin
coefficients are sensitive to new physics and poorly tested
experimentally, we analyze the possibility to extract them from a
future precision measurement of $\tau \lt(B_s \rt)/\tau\lt(B_d\rt)$.
Anticipating progress in the determination of the hadronic parameters
$\varepsilon_1$, $\varepsilon_2$ and $f_{B_s}/f_{B_d}$ we find that
the coefficient $C_4$ can be extracted with an uncertainty of order
$|\Delta C_4|\simeq 0.1$ from the double ratio
$(\tau(B_s)-\tau(B_d))/(\tau(B^+)-\tau(B_d))$, if
$|\varepsilon_1-\varepsilon_2|$ is not too small.

\thispagestyle{empty}
\newpage
\section{Introduction}
The theoretical achievement of the Heavy Quark Expansion (HQE)
\cite{hqe1} has helped a lot to understand the inclusive properties of
B-mesons.  The measurements of lifetime differences among the
b-flavoured hadrons test the HQE at the order $(\Lambda_{QCD}/m_b)^3$.
Today's experimental information on the B-meson lifetimes is in
agreement with the expectations from the HQE, but the present
theoretical predictions still depend on 4 poorly known hadronic
parameters $B_1$, $B_2$, $\varepsilon_1$ and $\varepsilon_2$
\cite{ns,bbd}. Recently they have been obtained by QCD sum rules
\cite{blls}.  Lattice results are expected soon from the Rome group
\cite{m} and will allow for significantly improved theoretical
predictions of the lifetime ratios.

Weak decays are triggered by a hamiltonian of the form 
\begin{eqnarray} 
H &=& \frac{G_F}{\sqrt{2}} \lt[ V_{\mathit{CKM}} \sum_{j=1}^2 C_j Q_j -
       V_{\mathit{CKM}}^\prime \lt( \sum_{k=3}^6 C_j Q_j + C_8 Q_8 \rt) \rt]. 
\label{hint}
\end{eqnarray}
Here $Q_1$ and $Q_2$ are the familiar current-current operators, $Q_3
\dots Q_6$ are penguin operators and $Q_8$ is the chromomagnetic
operator. Their precise definition is given below in \eq{basis}.  The
factors $V_{\mathit{CKM}}$ and $V_{\mathit{CKM}}^\prime$ represent the
factors stemming from the Cabibbo-Kobayashi-Maskawa matrix and are
specific to the flavour structure of the decay. Feynman diagrams in
which the spectator quark participates in the weak decay amplitude
induce differences among the various b-flavoured hadrons. Such
non-spectator effects have been addressed first by Bigi et al.\ in
\cite{spec} evaluating the matrix elements in the factorization
approximation in which $\varepsilon_1=\varepsilon_2=0$. Then Neubert
and Sachrajda \cite{ns} have found that even small deviations of
$\varepsilon_1,\varepsilon_2$ from zero drastically weaken the
prediction of \cite{spec} for the lifetime ratio $\tau (B^+)/\tau(B_d)
$, which can sizeably differ from 1.  On the other hand the deviation
of $\tau (B_s)/\tau(B_d) $ from unity has been estimated to be below
1\% in \cite{spec,ns} and the detailed analysis of Beneke, Buchalla
and Dunietz \cite{bbd}. Here $\tau (B_s)$ is the average lifetime of
the two CP-eigenstates of $B_s$.

Experimentally the ratio $\tau (B_s)/\tau(B_d) $ can also be addressed
by the measurements of the corresponding semileptonic branching
fractions. Since spectator effects in the semileptonic decay rate are
negligible, one may use $\tau (B_s)/\tau(B_d)=B_{SL}(B_s)/B_{SL}(B_d)$.  

So far only the effect of $Q_1$ and $Q_2$ has been considered in
\cite{spec,ns,bbd}. Taking into account the present experimental
uncertainty and the fact that $C_1$ and $C_2$ are much larger than
$C_{3-8}$ in the Standard Model this is justified. Yet once the
lifetime ratio $\tau (B_s)/\tau(B_d)$ is measured to an accuracy of a
few permille, the situation will change: The smallness of
$|\tau(B_s)/\tau(B_d)-1|$ is caused by the fact that the \textit{weak
  annihilation} contribution of $Q_{1,2}$ depicted in \fig{fig:cc}
almost yields the same contribution to the decay rates of $B_s$ and
$B_d$. The difference in the CKM-factors is negligible and the
lifetime difference is induced by the small difference of the
$(c,\ov{c})$ vs.\ $(c,\ov{u})$ phase space and by $SU(3)_F$ violations
of the hadronic parameters. These effects suppress
$|\tau(B_s)/\tau(B_d)-1|$ by roughly an order of magnitude compared to
$|\tau(B^+)/\tau(B_d)-1|$. The contributions stemming from the penguin
operators and the chromomagnetic operator, however, do not exhibit
such a cancellation. Their contribution to the non-spectator rate of
$B_s$ comes with the same power of the Wolfenstein parameter
$\lambda=0.22$ as the contribution of $Q_{1,2}$. In contrast the
effects of $Q_{3-8}$ to the non-spectator rate of $B_d$ or $B^+$ are
suppressed by two powers of $\lambda$ and are therefore negligible.
Hence one expects the contributions of $Q_{3-6}$ and $Q_8$ to
$|\tau(B_s)/\tau(B_d)-1|$ to be of the same order as those of $Q_1$
and $Q_2$. $\tau(B^+)/\tau(B_d)$ is not modified, so that the
phenomenological conclusions drawn from this ratio in \cite{ns} are
unchanged. Observables sensitive to $C_{3-8}$ like
$\tau(B_s)/\tau(B_d)$ are phenomenologically highly welcome. The
smallness of $C_{3-8}$ is a special feature of the helicity structure
of the corresponding diagrams in the Standard Model. In many of its
extensions the values of these coefficients can easily be much larger.
Such an enhancement due to supersymmetric contributions has been
discussed in \cite{k}. Up to now the focus of the search for new
physics has been on new contributions to $C_8$ \cite{k}. Yet many
interesting possible non-standard effects modify $C_{3-6}$ rather than
$C_8$: New heavy particles mediating FCNC at tree-level or
modifications of the $b$-$s$-$g$ chromoelectric formfactor affect
$C_{3-6}$, but not $C_8$.  Likewise new heavy coloured particles yield
extra contributions to $C_{3-6}$, e.g.\ in supersymmetry box diagrams
with gluinos modify $C_{3-6}$.

It is especially difficult to gain experimental information on the
numerical values of the penguin coefficients $C_{3-6}$. Even
penguin-induced decays to final states solely made of $d$ and $s$
quarks do not provide a clean environment to extract $C_{3-6}$: Any
such decay also receives sizeable contributions from $Q_2$ via
CKM-unsuppressed loop contributions \cite{cfms,lno}. In exclusive
decay rates these ``charming penguins'' preclude the clean extraction
of the effects of penguin operators \cite{cfms}. In semi-inclusive
decay rates like $B\rightarrow X_s \Phi$ the situation is expected
to be similar. In inclusive decay rates such as the total charmless
$b$ decay rate the effect of ``charming penguins'' can be reliably
calculated in perturbation theory. Yet these rates are much more
sensitive to new physics contributions in $C_8$ rather than in
$C_{3-6}$, because $Q_8$ triggers the two-body decay $b \rightarrow s
\,g$, while the effects of $Q_{3-6}$ involve an integration over
three-body phase space \cite{lno}. Notice from \fig{fig:cp} and 
\fig{fig:8}, however, that this phase space suppression of the terms 
involving $C_{3-6}$ is absent in the non-spectator diagrams 
inducing the lifetime differences.

\begin{nfigure}{tb}
\begin{minipage}[t]{0.48\textwidth}
\centerline{\epsfxsize=0.8\textwidth \epsffile{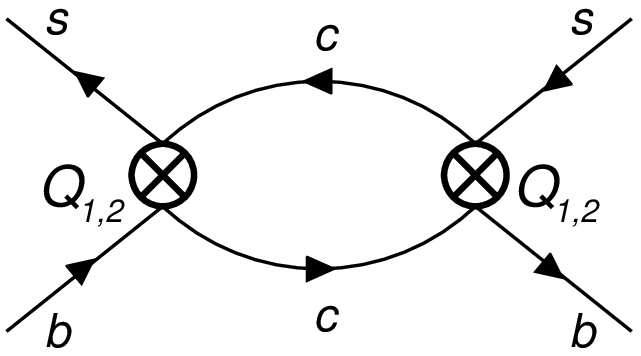}}
\caption{Non-spectator \emph{(weak annihilation)} contribution to the 
  $B_s$ decay rate involving two current-current operators. The
  corresponding diagram for the $B_d$ decay is obtained by replacing
  $s$ by $d$ and the upper $c$ by $u$.}\label{fig:cc}
\end{minipage}\hspace{2ex}
\begin{minipage}[t]{0.48\textwidth}
\centerline{\epsfxsize=0.8\textwidth \epsffile{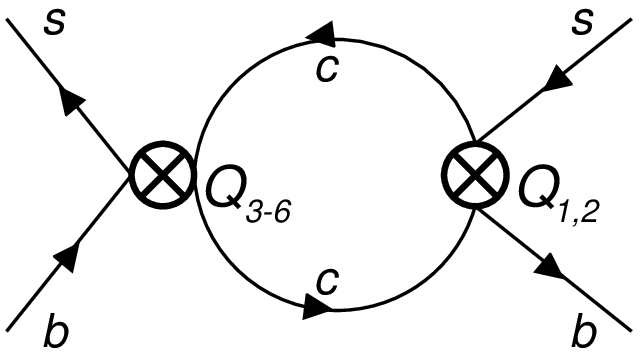}}
\caption{Weak annihilation diagram involving one penguin operator 
$Q_{3-6}$. Penguin contributions to the non-spectator rate of the 
$B_d$ meson are CKM suppressed and therefore negligible.  
}\label{fig:cp}
\end{minipage}
\end{nfigure}

This work is organized as follows: In the following section we
calculate the contributions to $\tau(B_s)/\tau(B_d)$ involving 
$Q_{3-6}$ or $Q_8$. Here we also obtain the dominant part of 
the radiative corrections to order $\alpha_s$. In sect.~\ref{sect:num} 
we discuss the phenomenological consequences within the Standard Model
and with respect to a potential enhancement of $C_{3-8}$ by new
physics. 

\section{Penguin Contributions}
For the non-spectator contributions to the $B_s$ decay rate we need
the $|\Delta B|=|\Delta S|=1$-hamiltonian:
\begin{eqnarray}
  H &=& \frac{G_F}{\sqrt{2}} V_{cb} V_{cs}^* 
        \left[ \sum_{j=1}^6 C_j Q_j + C_8 Q_8 \right] \; 
\label{hd}
\end{eqnarray}
with 
\begin{eqnarray}
Q_1 \; = \; (\bar{s}c)_{V-A}  \cdot (\bar{c}b)_{V-A} \cdot \tw ,
&& \qquad
Q_2 \; = \; (\bar{s}c)_{V-A}  \cdot (\bar{c}b)_{V-A} \cdot \1 \no \\[2mm]
Q_3 \; = \; \!\!\!\!\!\!\!
\sum_{q=u,d,s,c,b} (\bar{s}b)_{V-A}  \cdot (\bar{q}q)_{V-A}
  \cdot \1   ,
 && \qquad Q_4 \; = \; \!\!\!\!\!\!\! 
         \sum_{q=u,d,s,c,b} (\bar{s}b)_{V-A}  
                       \cdot (\bar{q}q)_{V-A} \cdot \tw 
\no \\[2mm] 
Q_5 \; = \; \!\!\!\!\!\!\!
 \sum_{q=u,d,s,c,b} (\bar{s}b)_{V-A}  \cdot (\bar{q}q)_{V+A}
  \cdot \1 ,
 && \qquad Q_6 \; = \; \!\!\!\!\!\!\! \sum_{q=u,d,s,c,b} 
(\bar{s}b)_{V-A}  \cdot (\bar{q}q)_{V+A}
  \cdot \tw  \no \\[2mm]
 Q_8 \; =\;  - \frac{g}{8 \pi^2} \, m_b \,\bar{s} \sigma^{\mu \nu} 
                 \lt( 1+\gamma_5 \rt)  T^a b \cdot G^a_{\mu \nu} 
 \; . \hspace{-6em}
\label{basis}
\end{eqnarray}
The colour singlet and non-singlet structure are indicated by \1\ and
\tw\ and $V\pm A$ is the Dirac structure. For more details see
\cite{lno,bjlw}. In \eq{hd} we have set $V_{ub} V_{us}^* =
O(\lambda^4)$ to zero.  The diagram of \fig{fig:cc} has been
calculated in \cite{ns,bbd} and yields contributions to the
non-spectator part $\Gamma^{\mathrm{non-spec}}$ of the $B_s$ decay
rate proportional to $C_2^2, C_1 \cdot C_2$ and $C_1^2$. The result
involves four hadronic matrix elements, which are parametrized by the
B-factors $B_1$, $B_2$, $\varepsilon_1$ and $\varepsilon_2$ \cite{ns}:
\begin{eqnarray}
\bra{B_s}  \ov{s} \gamma_{\mu} \lt( 1-\gamma_5 \rt) b \, 
           \ov{b} \gamma^{\mu} \lt( 1-\gamma_5 \rt) s   \ket{B_s} 
 & = & f_{B_s}^2 \, M_{B_s}^2 B_1 \nn 
\bra{B_s} \ov{s} \lt( 1+\gamma_5 \rt) b \,
          \ov{b} \lt( 1-\gamma_5 \rt) s \ket{B_s} 
 & = & f_{B_s}^2 \, M_{B_s}^2 B_2 \nn
\bra{B_s} \ov{s} \gamma_{\mu} \lt( 1-\gamma_5 \rt) T^a b \, 
          \ov{b} \gamma^{\mu} \lt( 1-\gamma_5 \rt) T^a s \ket{B_s} 
 & = & f_{B_s}^2 \, M_{B_s}^2 \varepsilon_1 \nn
\bra{B_s} \ov{s} \lt( 1+\gamma_5 \rt) T^a b \,
          \ov{b} \lt( 1-\gamma_5 \rt) T^a s \ket{B_s} 
 & = & f_{B_s}^2 \, M_{B_s}^2 \varepsilon_2 .
\end{eqnarray}
Here $T^a$ is the colour $SU(3)$ generator, $M_{B_s}=5369\pm 2$ MeV and
$f_{B_s}$ are the mass and decay constant of the $B_s$ meson.
$\tau(B_s)/\tau(B_d)-1$ is proportional to $\Gamma^{\mathrm{non-spec}}
(B_d)-\Gamma^{\mathrm{non-spec}} (B_s)$.  The main differences between
the result of \fig{fig:cc} for these two rates are due to the
different mass of $u$ and $c$ and the difference between $f_{B_d}$ and
$f_{B_s}$. Hence the current-current parts of $\tau(B_s)/\tau(B_d)-1$
proportional to $C_2^2, C_1 \cdot C_2$ or $C_1^2$ are suppressed by a
factor of $z$ or $\Delta$ with
\begin{eqnarray}
z \; = \; \frac{m_c^2}{m_b^2} \; = \; 0.085 \pm 0.023 ,
\qquad 
\Delta \; = \; 1- \frac{ f_{B_d}^2 M_{B_d} }{ f_{B_s}^2 M_{B_s} } 
\; = \; 0.23 \pm 0.11 . \label{latt}
\end{eqnarray}
The result for $\Delta$ in \eq{latt} is the present world average of
lattice calculations \cite{l}. There are also $SU(3)_F$ violations in
the B-factors, but they are expected to be small from the experience
with those appearing in $B^0-\ov{B}{}^0$-mixing. We want to achieve an
accuracy of 2 permille in our prediction for $\tau(B_s)/\tau(B_d)$,
which corresponds to an accuracy of 20-30\% in
$\tau(B_s)/\tau(B_d)-1$.  Therefore we use the same $B_1$, $B_2$,
$\varepsilon_1$ and $\varepsilon_2$ in $\tau(B_s)$ and $\tau(B_d)$.
Likewise there is $SU(3)_F$-breaking in the matrix elements of the
b-quark kinetic energy operator and the chromomagnetic moment
operator. These effects are suppressed by a factor of $m_b/(
\Lambda_{QCD} \cdot 16 \pi^2)$ with respect to those discussed above.
In \cite{bbd} they have been estimated from heavy meson spectroscopy
to be an effect of order one permille in $\tau(B_s)/\tau(B_d)$.

We are now interested in the diagram of \fig{fig:cp} involving one
large coefficient $C_{1,2}$ and one small penguin coefficient
$C_{3-6}$.  Diagrams with two insertions of penguin operators yield
smaller contributions proportional to $C_{3-6}^2$ and are neglected
here. To order $\lambda^2$ in $H$ we have $V_{\mathit{CKM}}^\prime =
0$ in \eq{hint} for the $B_d$ system and penguin effects are only
relevant in $\tau(B_s)$. Hence the penguin contributions to
$\tau(B_s)/\tau(B_d)-1$ do not suffer from the suppression factors $z$
and $\Delta$.  Next we want to evaluate the diagram of \fig{fig:8}
which encodes the interference of $Q_{1,2}$ with the chromagnetic
operator $Q_8$. This part of $\Gamma^{\mathrm{non-spec}}$ already
belongs to the order $\alpha_s$ and is small in the Standard Model,
but it can be sizeable in the new physics scenarios discussed in \cite{k}.

\begin{nfigure}{tb}
\begin{minipage}[t]{0.48\textwidth}
\centerline{\epsfxsize=0.9\textwidth \epsffile{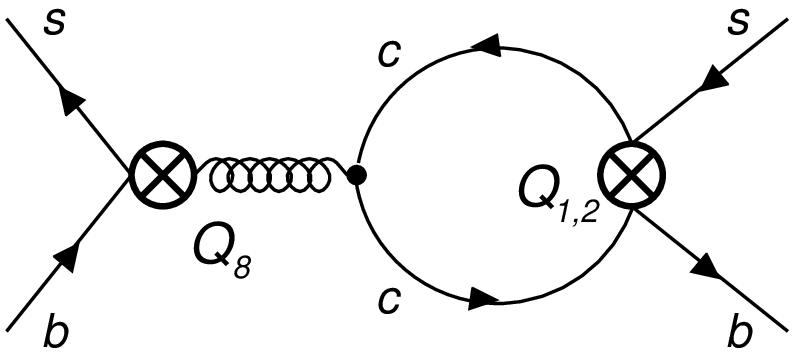}}
\caption{Contribution of $Q_8$ to $\Gamma^{\mathrm{non-spec}}(B_s)$. 
In the Standard Model the diagram is of the same order of magnitude 
as radiative corrections to \fig{fig:cp} and therefore negligible. 
Yet in models in which quark helicity flips occur in  flavour-changing 
vertices $|C_8|$ can easily be ten times larger than in the Standard 
Model \cite{k}. The contribution of $Q_1$ vanishes.}\label{fig:8}
\end{minipage}\hspace{2ex}
\begin{minipage}[t]{0.48\textwidth}
\centerline{\epsfxsize=0.9\textwidth \epsffile{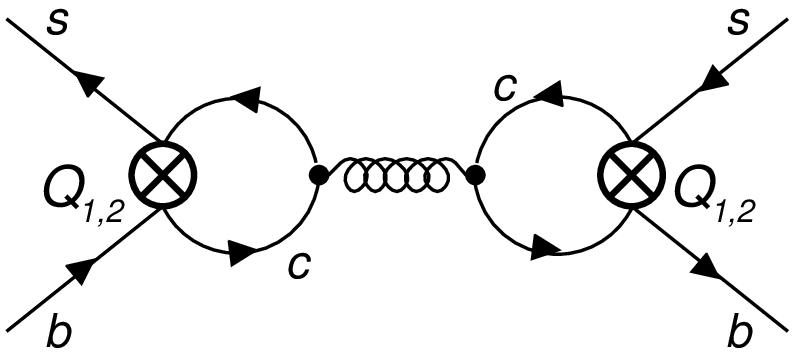}}
\caption{Penguin diagram contribution to
  $\Gamma^{\mathrm{non-spec}}(B_s)$. The final state corresponds
  to a cut through either of the $(\ov{c},c)$-loops. The contributions
  of $Q_1$ vanish by colour. This is the only NLO contribution to 
  $\tau(B_s)/\tau(B_d)-1$ involving $Q_{1,2}$ without suppression 
  factors of $\Delta$ or $z$. }\label{fig:nlo}
\end{minipage}
\end{nfigure}

We also must discuss radiative corrections to the contributions
involving the large coefficients $C_1$ and $C_2$. Dressing the diagram
in \fig{fig:cc} with gluons gives contributions to 
$\Gamma^{\mathrm{non-spec}}$ for both $B_d$ and $B_s$ and therefore 
yield small corrections of order $C_2^2 \Delta \alpha_s/\pi$ or less.  
The penguin diagram of \fig{fig:nlo}, however, contributes only to 
$\Gamma^{\mathrm{non-spec}}(B_s)$ in the order $\lambda^4$. Hence 
\fig{fig:nlo} yields an unsuppressed contribution of order 
$C_2^2 \alpha_s/\pi$ to $\tau(B_s)/\tau(B_d)-1$ and cannot be neglected. 
The result of these penguin loop diagrams can easily be absorbed into
the penguin coefficients $C_{3-6}$: In the result of the diagram of 
\fig{fig:cp} one must simply replace $C_{j}$ by 
\begin{eqnarray} 
C_j^\prime &=& C_j^{NLO} + \frac{\alpha_s}{4 \pi} C_2 \real \lt[ r_{2j} 
               \lt( 1, \sqrt{z}, \mu/m_b \rt) \rt] , 
               \qquad j=3,\dots,6 . \label{cprime}
\end{eqnarray} 
Here $r_{2j}$ encodes the result of the penguin diagram and can be
found in \cite{lno} in the NDR scheme. To cancel the scheme dependence
of $r$ we must also include the next-to-leading order (NLO)
corrections to $C_j$ as indicated in \eq{cprime}. More precisely: We
must include the NLO mixing of $C_2$ into $C_j$ in $C_j^{NLO}$,
$j=3,\dots,6$, but the penguin-penguin mixing only to the LO.  The
difference between these partial NLO-coefficients, which are tabulated
in \cite{lno}, and the full $C_j^{NLO}$'s has a negligible impact on
our result. Here we bypass this technical aspect of scheme
independence by tabulating the $C_j^\prime$'s in \tab{tab:wc}.
 
\begin{ntable}{tb}  
\begin{tabular}{l||r|r|r|r|r|r|r}
j & 1&2&3&4&5&6&8 \\[0.4ex]\hline
$C_j^{(0)} (\mu=m_b)$ & -0.249 & 1.108 & 0.011 & -0.026 & 0.008 &
                        -0.031 & -0.149 \\[0.2ex]
$C_j^{\prime} (\mu=m_b)$ &  &    & 0.014 &  -0.041 & 0.014 
                      & -0.047 & \\[0.4ex]\hline 
$C_j^{(0)} (\mu=m_b/2)$ & -0.361 & 1.169 &  0.017 &  -0.036 &  0.010 &
                        -0.048 & -0.166 \\[0.2ex] 
$C_j^{\prime} (\mu=m_b/2)$ &  &  &   0.017 &  -0.045 &
                             0.015 &    -0.058 &  \\[0.4ex]\hline
$C_j^{(0)} ( \mu=2 \,m_b)$ &  -0.167 &  1.067  &  0.007  &  -0.018 & 0.005
                        & -0.020 & -0.135 \\[0.2ex]
$C_j^{\prime} (\mu=2 \,m_b)$ &  & &  0.012 &  -0.036 &  0.013
                        &  -0.039 & 
\end{tabular}
\caption{The effective Wilson coefficients $C_j^\prime$ 
defined in \eq{cprime} for $z=0.085$, $\alpha_s(M_Z)=0.118$ and 
$m_b=4.8$ GeV. Varying $z$ within the range 
given in \eq{latt} affects the $C_j^\prime$ by 3-4 \% and is
negligible for our purposes. The $C_j^{(0)}$'s are the LO Wilson
coefficients.}\label{tab:wc}
\end{ntable}  

Our result for the non-spectator part of the $B_s$ decay rate 
reads:
\begin{eqnarray}
\Gamma^{\mathrm{non-spec}} \lt( B_s\rt)&=& - \frac{G_F^2 m_b^2}{12 \pi} 
        \lt| V_{cb} V_{cs} \rt|^2 \sqrt{1- 4 z} f_{B_s}^2 M_{B_s} 
        \lt[ a_1 \, \varepsilon_1 + a_2 \, \varepsilon_2 + 
             b_1 B_1 + b_2 B_2 \rt] \label{dgbs}
\end{eqnarray}
with 
\begin{eqnarray}
a_1 &=& \lt[ 2 C_2^2 + 4 C_2 C_4^\prime  \rt] \lt[1 - z \rt] + 
        12 z C_2 C_6^\prime + \lt[ 1 + 2 z  \rt] \frac{\alpha_s}{\pi} 
        C_2 C_8 \nn 
a_2 &=& - \lt[ 1 + 2 z \rt] \lt[ 2 C_2^2 + 4 C_2 C_4^\prime  +
               \frac{\alpha_s}{\pi} C_2 C_8 \rt] \nn 
b_1 &=&  \lt[ C_2 + N_c C_1  \rt]
        \lt\{ \lt( 1 - z  \rt) 
     \left[  \frac{C_2}{N_c}  +  C_1   + 
              2 C_3^\prime + 2 \frac{C_4^\prime}{N_c} \right]
                    + 6 z \lt[ C_5^\prime + \frac{C_6^\prime}{N_c} \rt] 
        \rt\} \nn
b_2 &=& - \lt[ 1 + 2 z  \rt] \lt[ C_2 + N_c C_1 \rt] \lt\{ 
        \frac{1}{N_c} \lt[ C_2 + N_c C_1 \rt] + 
      2 \lt[ C_3^\prime + \frac{C_4^\prime}{N_c} \rt]
        \rt\} \label{spc}
\end{eqnarray}
Here $N_c=3$ is the number of colours. By setting $C_j^\prime$,
$j=3,\dots,6$, and $C_8$ in \eq{spc} to zero one recovers the result
of \cite{ns}.\footnote{Notice that our notation of $C_1$ and $C_2$ is
  opposite to the one in \cite{ns}.}
The result for the non-spectator contributions to the $B_d$ decay
rate reads \cite{ns}:
\begin{eqnarray}
\!\!\!\!\!
\Gamma^{\mathrm{non-spec}} \lt( B_d\rt) \!\!\!
&=& \!\!\! \frac{G_F^2 m_b^2}{12 \pi} 
        \lt| V_{cb} V_{ud} \rt|^2 \lt(1- z\rt)^2 f_{B_s}^2 M_{B_s} 
        \lt( \Delta -1 \rt) 
        \lt[ a_1^d \, \varepsilon_1 + a_2^d \, \varepsilon_2 + 
             b_1^d B_1 + b_2^d B_2 \rt] \label{nsd}
\end{eqnarray}
with\footnote{In the large $N_c$ limit one finds
  $\Gamma^{\mathrm{non-spec}}$ helicity suppressed in analogy to the
  leptonic decay rate. This shows that one cannot neglect the
  $O(1/N_c)$ terms.}
\begin{eqnarray}
&&
\begin{array}[b]{rclrcl}
\displaystyle
a_1^d &=& \displaystyle 2 C_2^2 \lt( 1+ \frac{z}{2} \rt), \qquad &
\displaystyle
a_2^d &=& \displaystyle - 2 C_2^2 \lt( 1 + 2 z  \rt), \\
\displaystyle
b_1^d &=& \displaystyle \frac{1}{N_c} \lt( C_2 + N_c C_1 \rt)^2  
            \lt( 1 + \frac{z}{2} \rt), \qquad &
\displaystyle
b_2^d &=& \displaystyle - \frac{1}{N_c}
          \lt( C_2 + N_c C_1 \rt)^2  \lt( 1 + 2 z \rt) . 
\end{array} 
\label{specd}
\end{eqnarray}
When we combine (\ref{dgbs}-\ref{specd}) in order to predict 
$\tau (B_s)/\tau (B_d) -1$: 
\begin{subeqnarray}
\frac{\tau (B_s)}{\tau (B_d)} -1 &=& 
    \frac{\Gamma^{\mathrm{non-spec}}(B_d) - 
          \Gamma^{\mathrm{non-spec}}(B_s) }{
          \Gamma^{\mathrm{total}} } 
+O (10^{-3})
\nn
&=& K (z) \cdot
   \lt\{ 
      \Delta \lt[ 2 C_2^2 \lt( \varepsilon_1-\varepsilon_2 \rt) + 
                   \frac{ \lt( C_2+N_c C_1 \rt)^2}{N_c}
                          \lt( B_1 - B_2 \rt)  
              \rt] \rt. \slabel{res1}\\
&& \phantom{ \quad \big \{ }
   - 3 C_2^2 z \varepsilon_1 - \frac{3}{2} \frac{\lt( C_2+N_c C_1
    \rt)^2}{N_c} z B_1 \slabel{res2}\\
&& \phantom{ \quad\big \{ }
   + \Delta \, z \lt[ C_2^2 \lt( \varepsilon_1-4 \varepsilon_2 \rt) + 
                   \frac{ \lt( C_2+N_c C_1 \rt)^2}{2 N_c}
                          \lt( B_1 - 4 B_2 \rt)  
              \rt] \slabel{res3}\\
&& \phantom{ \quad\big \{ }
+ \lt[ 4 C_2 C_4^\prime  
       + (1 + 2 z) \frac{\alpha_s}{\pi} C_2 C_8 \rt] 
     \lt( \varepsilon_1 - \varepsilon_2 \rt) \slabel{res4}\\
&& \phantom{ \quad\big \{ }
   + 2 \lt( C_2 + N_c C_1  \rt)
   \lt(  C_3^\prime + \frac{C_4^\prime}{N_c}  
    \rt) \lt( B_1 - B_2 \rt) \slabel{res5}   \\
&& \phantom{ \quad\big \{ }
   - 4 z \, C_2 C_4^\prime \lt( \varepsilon_1 +  2 \varepsilon_2\rt)
   + 12 z \,  C_2 C_6^\prime \, \varepsilon_1 \nn
&& \phantom{ \quad\big \{ } \lt.
  + 2 z \lt( C_2 + N_c C_1  \rt) 
   \lt[ - \lt( C_3^\prime + \frac{C_4^\prime}{N_c}  \rt)
        \lt( B_1 + 2 B_2 \rt) + 
    3 \lt(   C_5^\prime + \frac{C_6^\prime}{N_c}  \rt) B_1  \rt]
   \rt\} \nn 
&& \phantom{ \quad\big \{ }
   + O ( 2 \cdot 10^{-3} ) \label{res} \no
\end{subeqnarray}
Here $K(z)$ reads
\begin{eqnarray}
K(z) &=& \frac{16 \pi^2 \lt| V_{ud} \rt|^2 B_{SL}}{m_b^3 f_1 (z) 
   \lt[ 1 + \alpha_s(\mu)/(2 \pi) \,h_{SL} \lt( \sqrt{z} \rt)   \rt]}
        f_{B_s}^2 M_{B_s} \lt[1 - 2 z \rt] \label{defk} \\
&\simeq& \frac{0.060}{1-4 \lt( \sqrt{z} -0.3 \rt) } \lt( 1 - 2 z \rt) 
 \, \frac{B_{SL}}{0.105} \lt( \frac{4.8}{m_b} \rt)^3
   \lt( \frac{f_{B_s}}{190 \, \mathrm{MeV}} \rt) ^2  .
\label{kexp}
\end{eqnarray}
In \eq{defk} we have used the common trick to evaluate the total width
$\Gamma^{\mathrm{total}}$ in terms of the semileptonic rate and the
measured semileptonic branching ration $B_{SL}$ via
$\Gamma^{\mathrm{total}}=\Gamma_{SL}/B_{SL}$. $f_1$ and $h_{SL}$ are
the phase space and QCD correction factor of $\Gamma_{SL}$ calculated
in \cite{nir}. We use the notation of \cite{lno}.  The approximation
in \eq{kexp} reproduces $K(z)$ to an accuracy of 3\%. The numerical
value of $h_{SL}$ entering \eq{kexp} corresponds to the use of the
one-loop pole mass ($\simeq 4.8$ GeV) for $m_b$.  For simplicity we
have expanded $K(z)$ and the terms in the curly braces in \eq{res} up
to the first order in $z$. The size of the error in \eq{res} is
estimated as $2 \cdot 10^{-3}$. Its main source is the
$SU(3)_F$-breaking in the kinetic energy and chromomagnetic moment
matrix elements appearing at order $\Lambda_{QCD}^2/m_b^2$ of the HQE,
which has been calculated to equal $(0$--$1)\cdot 10^{-3}$ in
\cite{bbd}.  Then terms of order $16 \pi^2 \Lambda_{QCD}^4/m_b^4$ can
maximally be of the same order of magnitude. Conversely the remaining
NLO correction of order $C_2^2 \, \Delta \, \alpha_s/\pi$ and the
CKM-suppressed contributions are much smaller. Likewise the
$SU(3)_F$-breaking in $\varepsilon_1,\varepsilon_2,B_1$ and $B_2$ is
expected to be at the level of a few percent and therefore smaller
than the present uncertainty in $\Delta$.

The first three lines (\ref{res1}-\ref{res3}) contain the result of 
the current-current operators calculated in \cite{ns,bbd}. The
remaining lines comprise the penguin effects. Note that the terms in 
(\ref{res4}-\ref{res5}) are neither suppressed by $\Delta$ nor by $z$. 
For $z=0$ the hadronic parameters in \eq{res} only appear in the
combinations $\varepsilon_1-\varepsilon_2$ and $B_1-B_2$, both of
which are of order $1/N_c$. The coefficients of $B_1-B_2$ suffer from  
numerical cancellations, e.g.\ $0.09 \leq C_2 + 3C_1 \leq 0.57$ 
(cf.\ \tab{tab:wc}), so that for  most values of the input parameters
only the terms involving $\varepsilon_1$ and $\varepsilon_2$ in
\eq{res1}, \eq{res2} and  \eq{res4} are important. 

Finally we discuss a potential systematic uncertainty: The derivation
of \eq{res} has assumed quark-hadron duality (QHD) for the sum over
the final states. QHD means that inclusive observables are unaffected
by the hadronization process of the quarks and gluons in the final
state.  The new results for inclusive observables in $B$ decays
presented at the 1997 summer conferences are consistent with QHD
\cite{n2}. There are two potential sources of QHD violation in our
problem: First it may be possible that the spectator decay rate of the
$b$-quark is affected by the hadronization process. Yet the ballpark
of this effect is independent of the flavour of the spectator quark
and cancels out in the ratio $\tau(B_s)/\tau(B_d)$. $SU(3)_F$-breaking
can only appear in the hadronization of the final state antiquark
which picks up the spectator quark and we do not expect the
$SU(3)_F$-breaking in the spectator decay rate to be larger than the
$SU(3)_F$-breaking in the $(\Lambda_{QCD}/m_b)^2$-terms of the HQE.
This effect should further not depend on whether the hadron containing
the spectator quark recoils against other hadrons or against a lepton
pair. Hence one can control the $SU(3)_F$-breaking in the spectator
decay rate by comparing the hadron energy in semileptonic $B_d$ and
$B_s$ decays.  More serious is a potential violation of QHD in the
non-spectator contribution $\Gamma^{\mathrm{non-spec}}$ itself. In a
theoretical analysis for the similar case of the width difference
$\Delta \Gamma_{B_s}$ of the two $B_s$ eigenstates the size of QHD
violation has been estimated to be moderate, maximally of order 30\%.
We can incorporate this into \eq{res} by assigning an additional error
of $\pm 0.3$ to $\Delta$.  In any case the issue of QHD violation in
lifetime differences will be experimentally tested in the forthcoming
years, when high precision measurements of $\tau(B^+)/\tau(B_d)$ and
of $\Delta \Gamma_{B_s}$ are confronted with accurate lattice results
for the hadronic parameters.
 
\section{Phenomenology}\label{sect:num}
In the following we want to investigate the numerical importance of
the penguin contribution. Then we analyze which accuracy is necessary
to detect or constrain new physics contributions to $C_{3-6}$ by 
a precision measurement of $\tau(B_s)/\tau(B_d)$. 

The three main hadronic parameters entering \eq{res} are $\Delta$, 
$f_{B_s}$ and $\varepsilon_1-\varepsilon_2$, while $B_1$ and 
$B_2$ come with small coefficients. The canonical sizes of the 
B-factors are $\varepsilon_i=O(1/N_c)$ and $B_i=1+O(1/N_c)$. An important
constraint on the $\varepsilon_i$'s is given by the measured value of 
$\tau(B^+)/\tau(B_d)$ \cite{ns}. The result of \cite{ns} for 
$\Gamma^{\mathrm{non-spec}} \lt( B^+ \rt) $ is obtained from 
\eq{nsd} by replacing the $a_i^d$, $b_i^d$'s with 
\begin{eqnarray}
\hspace{-3ex}
a_1^u &=& - 6 \lt( C_1^2+C_2^2 \rt), \qquad
b_1^u \; = \; - \frac{3}{N_c} \lt( C_2 + N_c C_1 \rt)^2 + 3 N_c C_1^2,
\qquad a_2^u \; = \; b_2^u \; = 0. \label{cfp}
\end{eqnarray}
The experimental world average \cite{f}
\begin{eqnarray}
\frac{\tau (B^+)}{\tau (B_d)} &=& 1.07 \pm 0.04 \label{expd}
\end{eqnarray}
leads to the following constraint:
\begin{eqnarray}
  \varepsilon_1 &\simeq & \lt( -0.2 \pm 0.1 \rt) \lt( \frac{0.17
  \,\mathrm{GeV}}{f_B} \rt)^2 \lt( \frac{m_b }{4.8 \,\mathrm{GeV}} \rt)^3 + 0.3
  \varepsilon_2 + 0.05 . \label{con}
\end{eqnarray}
In \cite{blls} the $\varepsilon_i$'s and $B_i$'s have been calculated
with QCD sum rules within the heavy quark effective theory (HQET). The
results are $\varepsilon_1 (\mu= m_b)=-0.08\pm 0.02$ and
$\varepsilon_2 (\mu= m_b)=-0.01\pm 0.03$ and $B_{1,2}=1+O(0.01)$. In
view of the smallness of the $\varepsilon_i$'s, however, it is
conceivable that other neglected effects are numerically relevant.For
example a NLO calculation of the matching between HQET and full QCD
amplitudes replaces $\varepsilon_i$ in \eq{dgbs} and \eq{nsd} by
$\varepsilon_i + d_i B_i$, where $d_i$ is a coefficient of order
$\alpha_s (m_b)/\pi$.  Here we will consider the range
$|\varepsilon_1|$, $|\varepsilon_2|\leq 0.3$, and further obey
\eq{expd}.

\begin{ntable}{tb}  
\begin{tabular}{r||*{3}{r|}|*{3}{r|}|*{3}{r|}|}
&
\multicolumn{3}{c||}{$\Delta=0.12$} &
\multicolumn{3}{c||}{$\Delta=0.23$} &
\multicolumn{3}{c||}{$\Delta=0.34$} \\\hline
$\varepsilon_1$ & -0.3 & -0.1 & 0 
              & -0.3 & -0.1 & 0  
              & -0.3 & -0.1 & 0  \\\hline\hline
$\varepsilon_2 = -0.3$ 
& 4.3 & 1.9 & *  & 4.3 & 4.5 & * & 4.3 & 7.1 & *
\\\hline
$\varepsilon_2 = -0.1$ 
& 3.7 & 1.3 & * & 1.1 & 1.3 & * & -1.5 & 1.3 & *
\\\hline
$\varepsilon_2 =  0.1$  
& * & 0.6 & -0.6 & -2.2 & -2.0  & -1.9 & -7.4 & -4.6 & -3.2
\\\hline 
$\varepsilon_2 =  0.3$ 
& * & -0.1 & -1.3 & * & -5.3 & -5.2 & -13.3 & -10.5 & -9.1
\end{tabular}
\caption{Standard Model prediction for $10^3\cdot 
\lt[\tau(B_s)/\tau(B_d)-1\rt]$ obtained
from \eq{res} for $f_{B_s}=190\, \mathrm{MeV}$, $\mu=m_b=4.8 \,
\mathrm{GeV}$, $z=0.085$, $\alpha_s(M_Z)=0.118$ and $B_1=B_2=1$.
The entries marked with * are in conflict with the experimental
constraint \eq{expd}, which also implies $\varepsilon_1 \lsim 0$.
There is an overall error of $\pm 2.0$ (see \ref{res}) for all 
entries. 
}\label{tab:num}
\end{ntable}

In \tab{tab:num} we have tabulated $\tau(B_s)/\tau(B_d)-1$ for various
values of $\Delta$ and $\varepsilon_1$, $\varepsilon_2$.  We
have further split $\tau(B_s)/\tau(B_d)-1$ into its current-current
part consisting of (\ref{res1}-\ref{res3}) and the new penguin part
involving $C_{3-6}^\prime,C_8$. These results can be found in \tab{tab:num2}.
\begin{ntable}{tb}  
\begin{tabular}{ r||*{5}{*{2}{r|}|} }
$\varepsilon_1-\varepsilon_2=$ & 
\multicolumn{2}{c||}{$-0.5$} &
\multicolumn{2}{c||}{$-0.3$} &
\multicolumn{2}{c||}{$-0.1$} &
\multicolumn{2}{c||}{$0.1$}  &
\multicolumn{2}{c||}{$0.2$} 
\\\hline
& peng & cc &  peng & cc &  peng & cc &  peng & cc &  peng & cc 
\\\hline\hline
$f_{B_s}=160\, \mathrm{MeV}$ & 
3.9 & -8.8 & 2.3 & -4.9 & 0.8 & -1.0 & -0.8 & 2.8 & -1.5 & 4.7 
\\\hline
$f_{B_s}=190\, \mathrm{MeV}$ & 
5.4 & -12.4 & 3.3 & -6.9 & 1.1 & -1.5 & -1.1 & 4.0 & -2.2 & 6.7
\\\hline
$f_{B_s}=220\, \mathrm{MeV}$ & 
7.3 & -16.6 & 4.4 & -9.3 & 1.5 & -2.0 & -1.4 & 5.3 & -2.9 & 9.0
\end{tabular}
\caption{The columns labeled with `peng' list the penguin contribution 
  to $10^3 \cdot\lt[\tau(B_s)/\tau(B_d)-1\rt]$ as a function of
  $\varepsilon_1-\varepsilon_2$ and $f_{B_s}$. The other input
  parameters have little impact on the size of the penguin
  contribution. The current-current part of $10^3
  \cdot\lt[\tau(B_s)/\tau(B_d)-1\rt]$ is listed for
  $\varepsilon_1=-0.1$ and $\Delta=0.23$. For the remaining parameters
  see \tab{tab:num}.  }\label{tab:num2}
\end{ntable}  
From \tab{tab:num2} we realize that the penguin contributions
calculated in this work are comparable in size, but opposite in sign 
to the current-current part obtained in \cite{bbd}. This makes the
experimental detection of any deviation of $\tau(B_s)/\tau(B_d)$ from
1 even more difficult, if the penguin coefficients are really
dominated by Standard Model physics. The results of \tab{tab:num} 
can be summarized as 
\begin{eqnarray}
\frac{\tau(B_s)}{\tau(B_d)} -1 &=& \lt( -1.2 \pm 8.0 \pm 2.0 \rt)
\cdot 10^{-3} \cdot \lt(\frac{f_{B_s}}{190 \, \mathrm{MeV}} \rt)^2
\lt( \frac{4.8 \,\mathrm{GeV}}{m_b } \rt)^3 . \label{nums}
\end{eqnarray}
Here the first error stems from the uncertainty in $\varepsilon_1$ and
$\varepsilon_2$ and will be reduced once lattice results for the
hadronic parameters are available. The second error summarizes the
remaining uncertainties. If $\Delta$ and $\varepsilon_2$
simultaneously aquire extreme values, $\tau(B_s)/\tau(B_d) -1$
can be slightly outside the range in \eq{nums} (see \tab{tab:num}).

Today we have little experimental information on the sizes of the
penguin coefficients. Their smallness in the Standard Model allows for
the possibility that they are dominated by new physics.  The total
charmless inclusive branching fraction $Br(B \rightarrow \textit{no
  charm})$ is a candidate to detect new physics contributions to $C_8$
\cite{k}, but it is much less sensitive to $C_{3-6}$ \cite{lno}. The
decreasing experimental upper bounds on $Br(B \rightarrow \textit{no
  charm})$ \cite{f} therefore constrain $C_8$ but leave room for a
sizeable enhancement of $C_{3-6}$. Now \eq{res} reveals that
$\tau(B_s)/\tau(B_d)$ is a complementary observable mainly sensitive
to $C_4 $, while $C_8$ is of minor importance. As mentioned in the
introduction, many interesting new physics scenarios affect $C_{3-6}$,
but not necessarily $C_8$. We remark here that we constrain ourself to
new physics scenarios, in which the CKM factors of the new
contributions are the same as the ones of the Standard Model. This is
fulfilled to a good approximation in most interesting models \cite{k}.
Now any new physics effect modifies $C_{3-6}$ at some high scale of
the order of the new particle masses, while the Wilson coefficients
entering \eq{res} are evaluated at a low scale $\mu \approx m_b$. The
renormalization group evolution down to $\mu \approx m_b$ mixes the
new contributions to $C_{3-6}$.  New physics contributions $\Delta
C_{3-6} ( \mu = 200 GeV) $ affect $C_4 ( \mu = 4.8\, \mathrm{GeV})$ by
\begin{eqnarray}
\Delta C_4 ( \mu = 4.8\, \mathrm{GeV}) &=& 
-0.35 \, \Delta C_3 ( 200 \, \mathrm{GeV}) + 
 0.99 \, \Delta C_4 ( 200  \, \mathrm{GeV})  \nn
&& -0.03 \, \Delta C_5 ( 200 \, \mathrm{GeV})  
 -0.22 \, \Delta C_6 ( 200 \,  \mathrm{GeV}) . \no
\end{eqnarray}   
Observe that $\Delta C_4 (200 \, \mathrm{GeV})=-0.05$ already increases
$C_4^\prime (m_b)$ by more than a factor of two. 

Clearly the usefulness of $\tau(B_s)/\tau(B_d)$ to probe $C_{3-6}$
crucially depends on the size of $|\varepsilon_1-\varepsilon_2|$ and
$f_{B_s}$. We now investigate the sensitivity of $\tau(B_s)/\tau(B_d)$
to $\Delta C_4 (\mu = m_b)$ in a possible future scenario for 
the hadronic parameters. We assume
\begin{eqnarray}
\varepsilon_1 \; = \; -0.10 \pm 0.05, && \qquad 
\varepsilon_2 \; = \; 0.20 \pm 0.05,  \qquad
B_1,B_2 \; = \; 1.0 \pm 0.1, \nn
f_{B_s} \; = \; \lt(190 \pm 15 \rt) \, \mathrm{GeV}, && \qquad 
\Delta \;  = \; 0.23 \pm 0.05, \qquad
m_b \; = \; \lt( 4.8 \pm 0.1 \rt) \, \mathrm{GeV}. \label{sc}
\end{eqnarray}
The assumed accuracy for $f_{B_s}$ will be achieved, once more
experimental information on the $B_s$ system is obtained, e.g.\ after
the detection of $B_s-\ov{B}_s$-mixing. Also a more precise
measurement of $f_{D_s}$ is helpful, because lattice QCD predicts the
ratio $f_{B_s}/f_{D_s}$ much better than $f_{B_s}$ \cite{l}. The error
bars of the other hadronic parameters likewise appear within reach, if
one keeps in mind that information on $\varepsilon_1$ and
$\varepsilon_2$ will not only be obtained from the lattice but also
from other observables like $\tau(B^+)/\tau(B_d)$. Experimental
progress in \eq{expd} and a next-to-leading order calculation of the
coefficients in \eq{cfp} and \eq{specd} will significantly improve the
constraint in \eq{con}. In \fig{fig:plot} we show the dependence of
$\tau(B_s)/\tau(B_d)-1$ on $\Delta C_4 (\mu)$ for the scenario in
\eq{sc}. A cleaner observable is the double ratio
\begin{eqnarray}
\frac{\tau(B_s)-\tau(B_d)}{\tau(B^+)-\tau(B_d)} &=& 
\frac{B_{SL}(B_s)-B_{SL}(B_d)}{B_{SL}(B^+)-B_{SL}(B_d)} ,
 \label{dr}
\end{eqnarray} 
which depends on $\varepsilon_1, \varepsilon_2$  and $\Delta$, while the 
dependence on $f_{B}$ and $m_b$ cancels. The corresponding plot for 
the parameter set of \eq{sc} can be found in \fig{fig:plot2}

\begin{nfigure}{tb}
\centerline{\epsfxsize=0.9\textwidth \epsffile{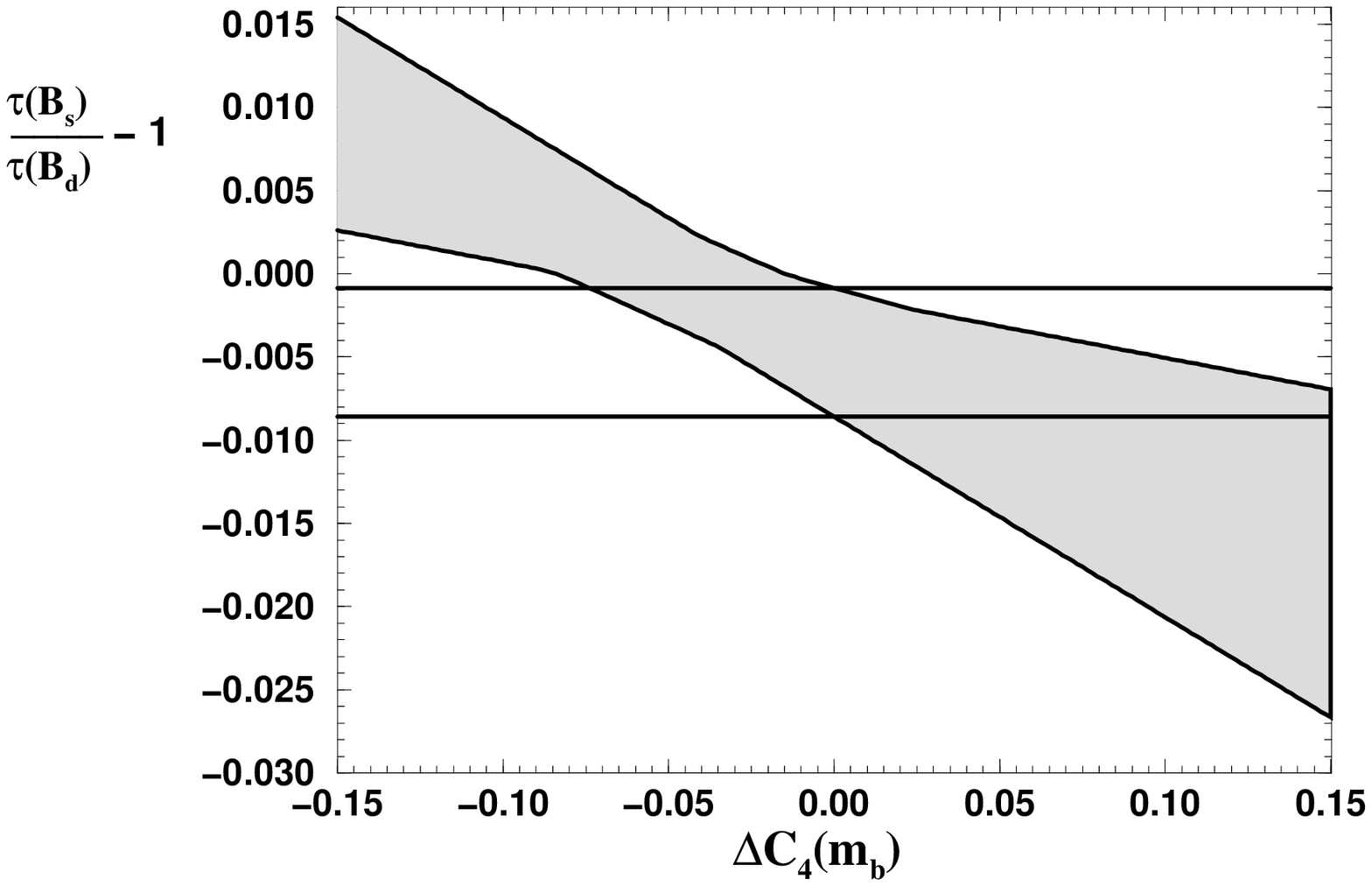}}
\caption{Dependence of $\tau(B_s)/\tau(B_d)-1$ on a new physics
  contribution $\Delta C_4$.  The shaded area corresponds to a
  variation of the input parameters within the range of \eq{sc}.  The
  horizontal lines mark the Standard Model range corresponding to
  $\Delta C_4=0$.  }\label{fig:plot}
\end{nfigure}
\begin{nfigure}{tb}
\centerline{\epsfxsize=0.9\textwidth \epsffile{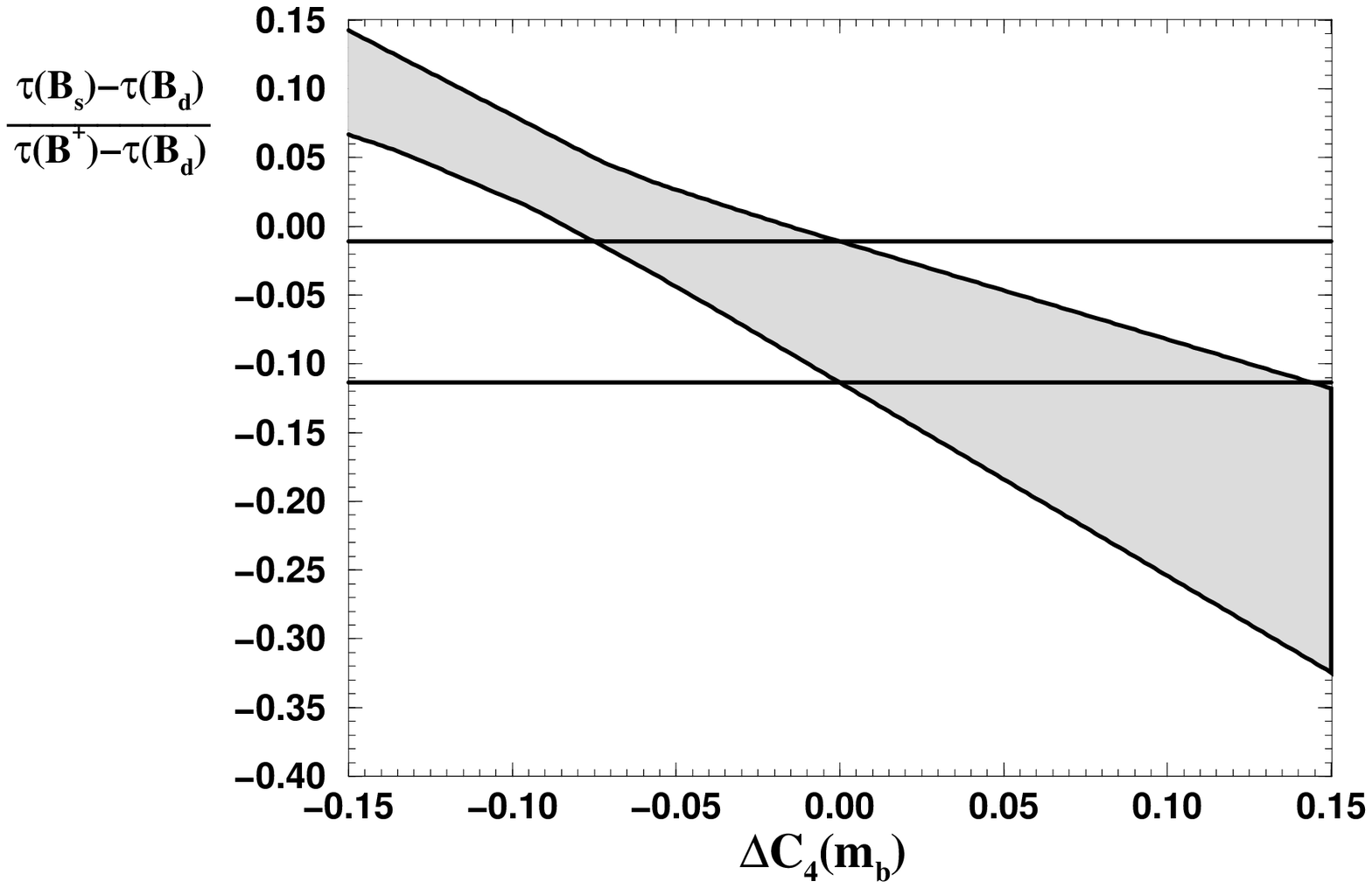}}
\caption{Dependence of $(\tau(B_s)-\tau(B_d))/(\tau(B^+)-\tau(B_d))$ 
  on $\Delta C_4$ for the parameter set in \eq{sc}. This double ratio
  depends on $f_{B_s}$ and $f_{B_d}$ only through $\Delta$, and the 
  factor of $m_b^{-3}$ in \eq{res} cancels. 
}\label{fig:plot2}
\end{nfigure}

We find a smaller error band for
$(\tau(B_s)-\tau(B_d))/(\tau(B^+)-\tau(B_d))$ than for
$\tau(B_s)/\tau(B_d)-1$. If $\Delta C_4 < -0.075$ or $\Delta C_4 >
0.140$, we find the allowed range for
$(\tau(B_s)-\tau(B_d))/(\tau(B^+)-\tau(B_d))$ incompatible with the
Standard Model. An experimental lower bound
$\tau(B_s)/\tau(B_d)>1.005$ would indicate a new physics contribution
$\Delta C_4<-0.063$ in our scenario. Likewise the experimental
detection of a sizeable negative lifetime difference
$\tau(B_s)-\tau(B_d)$ may reveal non-standard contributions to
$C_4^\prime$ of similar size as its Standard Model value.
\fig{fig:plot2} shows that e.g.\ the bound $\tau(B_s)-\tau(B_d)< -0.20
(\tau(B^+)-\tau(B_d))$ would indicate $\Delta C_4 > 0.051$.  We
conclude that the detection of new physics contributions to $C_4$ of
order $0.1$ is possible with precision measurements of
$\tau(B_s)/\tau(B_d)$. 
 
\section{Conclusions}
We have calculated the contributions of the penguin operators
$Q_{3-6}$, of the chromomagnetic operator $Q_8$ and of penguin
diagrams with insertions of $Q_2$ to the lifetime splitting between
the $B_s$ and $B_d$ meson. In the Standard Model the penguin effects
are found to be roughly half as big as the contributions from the
current-current operators $Q_1$ and $Q_2$, despite of the smallness of
the penguin coefficients. Yet they are opposite in sign, so that any
deviation of $\tau(B_s)-\tau(B_d)$ from zero is even harder to detect
experimentally.  Assuming a reasonable progress in the determination
of the hadronic parameters a precision measurement of
$\tau(B_s)/\tau(B_d)$ can be used to probe the coefficient $C_4$ with
an accuracy of $|\Delta C_4|=0.1$. Hence new physics can only be
detected, if $C_4$ is dominated by non-standard contributions.  The
sensitivity to $C_4$ depends crucially on the difference of the
hadronic parameters $\varepsilon_1$ and $\varepsilon_2$.  For the
extraction of $C_4$ the double ratio
$(\tau(B_s)-\tau(B_d))/(\tau(B^+)-\tau(B_d))$ turns out to be more
useful than $\tau(B_s)/\tau(B_d)$.

\section*{Acknowledgements}
U.N.\ is grateful for stimulating discussions with Martin Beneke,
Matthias Neubert and Chris Sachrajda. Y.Y.\ Keum thanks
Prof.\ W.~Buchm\"uller for his hospitality at DESY.

\end{document}